\documentclass{article}

\usepackage[english]{babel}

\usepackage[letterpaper,top=2cm,bottom=2cm,left=3cm,right=3cm,marginparwidth=1.75cm]{geometry}

\usepackage[utf8x]{inputenc}
\usepackage{amsmath}
\usepackage{graphicx}
\usepackage[colorlinks=true, allcolors=blue]{hyperref}
\usepackage{authblk}
\usepackage{floatrow}
\newfloatcommand{capbtabbox}{table}[][\FBwidth]

\title{AixBench: A Code Generation Benchmark Dataset}
\author[1]{Yiyang Hao}
\author[2]{Ge Li}
\author[1]{Yongqiang Liu}
\author[1]{Xiaowei Miao}
\author[1]{He Zong}
\author[1]{Siyuan Jiang}
\author[1]{Yang Liu}
\author[1]{He Wei}
\affil[1]{aiXcoder \\ \{haoyiyang, liuyongqiang, miaoxiaowei, zonghe, jiangsiyuan, liuyang, weihe\}@aixcoder.com}
\affil[2]{Peking University \\ lige@pku.edu.cn}

\begin{document}
\maketitle

\begin{abstract}
We present a benchmark dataset for evaluating method-level code generation task. The benchmark contains a dataset of 175 samples for automated evaluation and a dataset of 161 samples for manual evaluation. We also present a new metric for automatically evaluating the correctness of the generated code, and a set of criteria to manually evaluating the overall quality of the generated code.
\end{abstract}

\section{Introduction}

This is a method-level benchmark for evaluating code generation models, which take natural language as input and code as output, and is primarily used to evaluate the ability of code generation models. AixBench is divided into two datasets (see Table~\ref{tab:statistics}):

\begin{enumerate}
\item Automated Test Dataset

Each sample in this part of the dataset contains a functionally independent and well-described natural language function description, the function signature of the function, and a set of unit tests that verify the correctness of this function.

The main use of this dataset is to automatically evaluate the correctness of the code generated by the model.

\item NL Task Description Dataset

Each sample in this part of the data set contains a relatively independent functional description. This part of the data is closer to the real method description in the code, and contains some functional descriptions whose details are not very clear.

Please refer to \ref{evaluationStandard} for the human evaluation criteria.

\end{enumerate}

\begin{table}
\centering
\begin{tabular}{l|r|r}
Datasets & Automated Test Dataset & NL Task Description Dataset \\\hline
Test Set Size & 175 & 161
\end{tabular}
\caption{\label{tab:statistics}Data statistics of the two datasets of AixBench.}
\end{table}

Currently, these two datasets only contain Java codes, and the natural language description part contains English and Chinese languages. If you only care about code correctness, you can just use the automated test dataset.

\section{Backgrounds and Related Work}

Commonly used metrics, such as Exactly Match, BLEU or Perplexity, are not suited for evaluating the correctness of method-level code generation, because when considering the correctness of a program, the difference in many details between two pieces of generated code, like the name of a variable, the order of two data-flow independent instructions, the way loops and branches are written, and sometimes even the algorithm, do not directly decide which program is ``correct'' or not. In actual software development, people commonly use test cases to ensure a certain function works as intended. Therefore each sample of the Automated Test Dataset contains several hand-crafted automated test cases to ensure the correctness of the generated code.

The closest dataset to our Automated Test Dataset is HumanEval\cite{chen2021codex}, released by OpenAI together with Codex model. Their dataset contains 164 hand-written programming problems together with examples and test cases. However we find this insufficient to test our aiXcoder XL code generation model because of three reasons: One, their dataset is purely in Python and our model is fine-tuned on Java; Two, the problems are mostly about pure algorithm and string manipulation, which are only a small subset of real world problems. Three, the prompts in HumanEval contain examples, which is hardly written by human in a ``text-code'' interaction scenario.

Another dataset for the same purpose is APPS \cite{hendrycksapps2021}, which also includes description, examples, and test cases. APPS falls short for the same reasons as HumanEval: being in Python only and mostly contains algorithm and programming contests problems instead of challenges developers face in daily work.

An additional reason for us to create yet another automated code generation benchmark is that none of the datasets that we know of contains non-English prompts. And as non-native English speakers, we know well that how well a model adapts to the native language of the user affects users' accessibility by a lot.

PandasEval and NumpyEval\cite{cert} are similar to HumanEval but more limited to specific Python packages.

Other datasets like CONCODE\cite{concode} or PY150 do not include test cases at all.

\section{Task Description}

The goal of a code generation model is to generate a piece of code from a piece of natural language description. To automate the tests, we also add the signature of the desired function to the model. A signature of a function defines how this function should be called, by specifying the return type and the types and names and the order of the parameters. In this paper, we call the piece of natural language description from the input as ``task''.

\section{Datasets}

AixBench contains two datasets: Automated Test Dataset for mostly-automated code correctness evaluation and NL Task Description Dataset for manual code quality evaluation.

\subsection{Automated Test Dataset}

This data is a collection of hand-picked batches of ``Method Comments'' from open-sourced ``Method Comments - Method Implementation'' pairs. Our selection criteria are:

\begin{enumerate}
    \item Comments well describe a function that can be implemented.
    \item The functions are relatively independent and do not depend on the understanding of the context of the project and business logic.
    \item The functionality is reasonable and could occur in a developer's day-to-day work. rather than programming competition quizzes or coursework.
    \item Comments are descriptions of the objective, rather than descriptions of the implementation process.
\end{enumerate}

On this basis, we extracted the descriptions in the comments, and then made some supplements, so that:

\begin{enumerate}
    \item The description contains specific information necessary to implement the function. For example: \texttt{Returns whether or no the JDK version is high enough.} There is no clear ``high enough'' standard. So we added it manually as \texttt{Returns whether or no the JDK version is 1.7u40 and above.} This step is needed purely because we want to automate the tests.
    \item The irrelevant part of description is deleted. For example we removed the second half of the \texttt{max() that works on three integers. Like many of the other max() functions in this class.} from the original data.
\end{enumerate}

Natural language descriptions naturally will contain certain grammatical errors or punctuation or inconsistencies in capitalization. We keep these because we think these perturbations test the model's anti-disturbance ability.

Each sample in the Automated Test Dataset contains a \texttt{raw\_nl} and a \texttt{signature}. \texttt{raw\_nl} is the natural language description and \texttt{signature} is the name and the parameters of the desired function. \texttt{signature} exists solely because of the need of automating tests.

\begin{figure}
\centering
\includegraphics[width=1\textwidth]{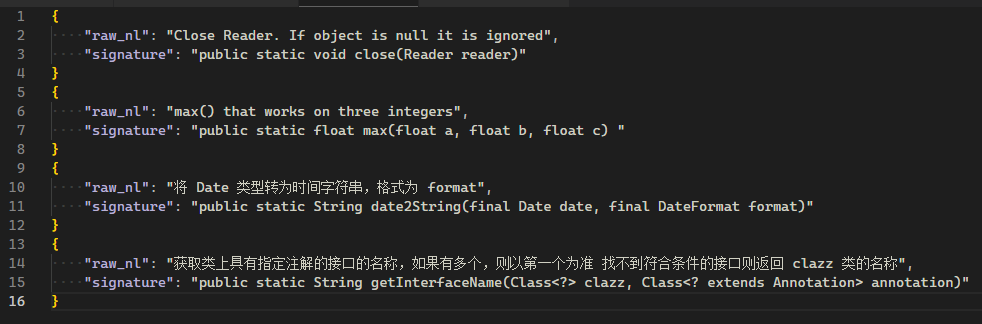}
\caption{\label{fig:autodata_example}Example data from the Automated Test Dataset. Test cases are implemented separately in source files.}
\end{figure}

\subsection{NL Task Description Dataset}

This data is a collection of hand-picked batches of ``Method Comments'' from open-sourced ``Method Comments - Method Implementation'' pairs. Our selection criteria are:

\begin{enumerate}
    \item Comments well describe a function that can be implemented.
    \item The functions are relatively independent and do not depend on the understanding of the context of the project and business logic.
    \item The functionality is reasonable and could occur in a developer's day-to-day work. rather than programming competition quizzes or coursework.
    \item We allow a certain degree of ambiguity, such as in \texttt{Read the encoded image data from a JPEG image.}, we do not specify how the read data should be handled. During evaluation, as long as the code generated by the model fully implements the functions described in the description, then a full score is awarded for correctness.
\end{enumerate}

\begin{figure}
\centering
\includegraphics[width=1\textwidth]{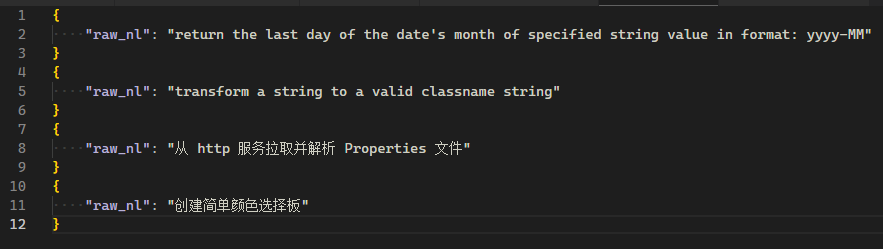}
\caption{\label{fig:manualdata_example}Example data from the NL Task Description Dataset.}
\end{figure}

\subsubsection{Evaluation Criteria}\label{evaluationStandard}

We manually evaluate the code generated by the model in three dimensions.

\begin{enumerate}
    \item Correctness:
    \begin{enumerate}
        \item 4 points: The specified function is fully realized.
        \item 3 points: The main function is realized. However, some details are missing, which does not affect the correctness of the overall logic. A little modification is need to meet all the requirements.
        \item 2 points: Only the core function is implemented. Most of the requirements are not reflected in the code. More modifications are required to meet the requirements.
        \item 1 point: The specified function is not implemented at all.
    \end{enumerate}
    \item Code Quality:
    \begin{enumerate}
        \item 3 points: The details are in place. No obviously better code in terms of performance exists. If possible, resources are released accordingly. No obvious code smell.
        \item 2 points: Some details are not in place. There is code smell of low severity.
        \item 1 point: There is significantly better solution in terms of performance. Or there is serious code smell.
    \end{enumerate}
    \item Maintainability:
    \begin{enumerate}
        \item 5 points: The method implementation is very standardized, the variable naming is semantically straightforward, the method is not unnecessarily bloated, the readability is good, the code is short, and the code blocks are clearly structured.
        \item 4 points: The method implementation is relatively standardized, the variable naming is basically semantically straightforward, and the readability is better.
        \item 3 points: The method implementation meets certain specifications, some variable names are meaningless, and defective code and deprecate methods are used.
        \item 2 points: The code is written in a confusing way, or does not follow a consistent specification, or there are many meaningless names in variable naming, or there are certain repetitions and redundant codes. Poor readability.
        \item 1 point: Very confusing, completely illogical, hard-to-read code.
    \end{enumerate}
\end{enumerate}

\section{Experiment}

In addition to pass@1, a special case of pass@k where k=1, introduced in \cite{chen2021codex}, we also use the average test cases pass ratio (AvgPassRatio) to evaluate the code generation model. $AvgPassRatio$ can be calculated like this:

\[AvgPassRatio = \frac{1}{n}\sum_{i}^{n} PassRatio_i\]
\[PassRatio_i = \frac{Count_{i,pass}}{Count_{i,total}} \]

where $Count_{i,pass}$ is the number of passed test cases in sample $i$ and $Count_{i,total}$ is the total number of test cases in sample $i$.

We prefer AvgPassRatio over pass@k because we want to directly measure how helpful the generated code can be for developers. Intuitively, a program that passes 90\% of the test cases is already good enough to help a developer implement that task, but this program will fail in pass@k because pass@k requires 100\% of test cases passed.

We evaluated aiXcoder XL\cite{aixcoderXL} and GitHub Copilot\cite{githubcopilot} on our datasets. The results show that both model perform similarly on both automatic tests and manual evaluation.

\begin{figure}
\begin{floatrow}
\capbtabbox{%
    \begin{tabular}{c|p{2cm}|p{2cm}}
         Metrics & aiXcoder XL & Copilot \\\hline
         pass@1 & 86\newline(49.14\%) & 81\newline(46.29\%) \\
         AvgPassRatio & 120.1979 (68.68\%) & 121.7152 (69.55\%)
    \end{tabular}
}{%
    \caption{Automatic Comparison on Correctness over 175 samples}%
}
\ffigbox{%
  \includegraphics[width=0.4\textwidth]{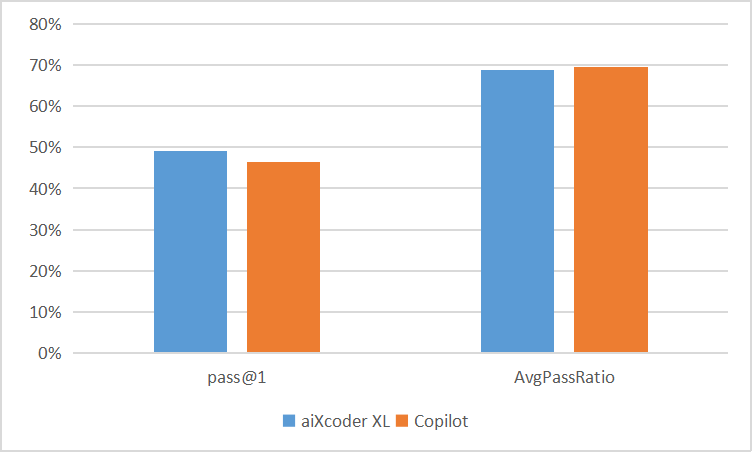}
}{%
  \caption{Automatic Comparison on Correctness}
}
\end{floatrow}
\end{figure}

\begin{figure}
\begin{floatrow}
\capbtabbox{%
    \begin{tabular}{c|p{2cm}|p{2cm}}
         Metrics & aiXcoder XL & Copilot \\\hline
         Correctness & 2.9503 & 2.9875 \\
         Code Quality & 2.2049 & 2.1739 \\
         Mantainability & 2.9937 & 3.0931
    \end{tabular}
}{%
    \caption{Manual Comparison between aiXcoder XL and Copilot on NL Task Description Dataset}
    \label{tab:manual_compare_table}
}
\ffigbox{%
    \includegraphics[width=0.4\textwidth]{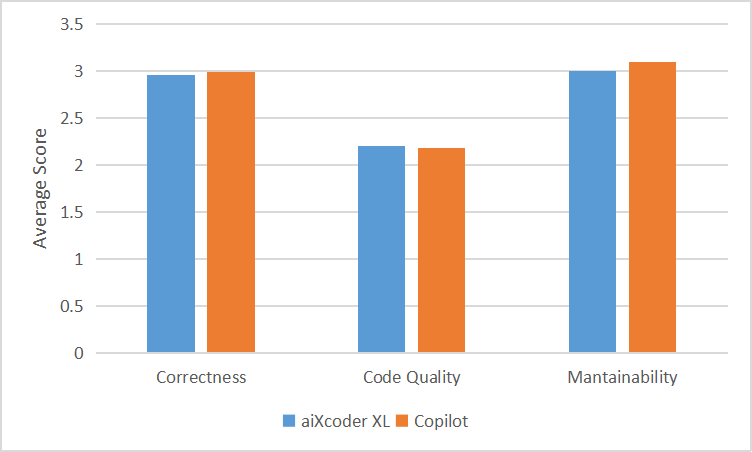}
}{%
    \caption{Manual Comparison between aiXcoder XL and Copilot on NL Task Description Dataset}
}
\end{floatrow}
\end{figure}

\section{Conclusion}

In this paper, we present a benchmark for automatically evaluating correctness and manually evaluating overall quality of the generated code for code generation models. And we also evaluated two released code generation products, aiXcoder XL and GitHub copilot on this benchmark.

\bibliographystyle{alpha}
\bibliography{main}

\end{document}